# MLS and CALIPSO Evidence for a Warm-Pool Transition in Tropical Convective Moistening and Cloud Ice

Isabel Nucera[1], Jonathan H. Jiang[2,3], Yiyang Zhang[4], King-Fai Li[5], Bin Zhao[6], Hwayon Choi[7], Yong-Sang Choi[7]

[1] Pacific Academy, Irvine, California, U.S.A.

[2] Jet Propulsion Laboratory (Retired), California Institute of Technology, Pasadena, California, USA

[3] Joint Institute for Regional Earth System Science & Engineering, University of California, Los Angeles, California, USA

[4] Department of Atmospheric and Oceanic Sciences, Fudan University, Shanghai, China

[5] Environmental Sciences Department, University of California, Riverside, California, USA

[6] State Key Laboratory of Regional Environment and Sustainability, School of Environment, Tsinghua University, Beijing, China

[7] Department of Climate and Energy Systems Engineering, Ewha Womans University, Seoul, South Korea

Corresponding: Jonathan H. Jiang, email: jonathanjiang@ucla.edu

## Abstract

Upper-tropospheric water vapor and cloud ice associated with tropical deep convection are central to water-vapor and high-cloud feedbacks. We investigate a warm-pool transition in tropical convective moistening using 20 years of Aura Microwave Limb Sounder (MLS) Version 5 $H_2O$ and cloud-ice observations (August 2004–December 2024), together with Cloud-Aerosol Lidar with Orthogonal Polarization (CALIOP) aboard the Cloud-Aerosol Lidar and Infrared Pathfinder Satellite Observations (CALIPSO) mission (June 2006–June 2023). The combined day-plus-night CALIPSO product (CALIPSO-A) provides an independent ice-cloud constraint; the $H_2O$ column shown on the CALIPSO grid is derived from Modern-Era Retrospective Analysis for Research and Applications, Version 2 (MERRA-2), not from CALIOP. We combine objective sea surface temperature (SST) breakpoint fitting, cross-sensor ice water path (IWP) comparisons, Clouds and the Earth's Radiant Energy System (CERES) longwave, shortwave, and net cloud radiative effect (CRE), and Integrated Multi-satellitE Retrievals for GPM (IMERG) precipitation diagnostics from the Global Precipitation Measurement (GPM) mission. MLS and CALIPSO-A show cloud-ice maxima over deep-convective warm-pool regions and identify warm-regime SST breakpoints near 299–300 K. Normalized MLS and CALIPSO-A IWP vary coherently in the western Pacific and tropical mean. CERES indicates that stronger longwave CRE in the warm regime is partly offset by stronger shortwave cooling, while GPM IMERG confirms that MLS IWP increases with convective rain rate. These results show that convective moistening and cloud-ice production across the tropical warm-pool transition are persistent cross-sensor features of the tropical atmosphere.

**Significance Statement**

Deep tropical thunderstorms inject water vapor and ice high into the atmosphere, where they strongly influence how Earth exchanges radiation with space. This study uses two independent satellite records, Aura MLS and CALIPSO/CALIOP, plus CERES radiation and GPM precipitation data, to show that tropical cloud ice and water vapor increase sharply in the warmest ocean regions near 299–300 K. The results demonstrate that SST-dependent convective moistening and condensate production persist across two decades and across observing systems. They also clarify that the observed IWP increase constrains convective moistening capacity, while a direct iris-effect test would require explicit normalization by convective activity or cloud area.

## 1. Introduction

Upper-tropospheric water vapor and tropical high clouds exert a disproportionate influence on Earth's radiation budget because small changes in humidity, ice condensate, cloud-top height, and cloud amount near the tropopause can substantially affect outgoing longwave radiation. The tropical warm pool is particularly important because it concentrates the deepest convection, strongest latent heating, and largest upper-tropospheric moistening. A robust observational constraint on this region must therefore address not only the mean pattern of moisture and ice, but also whether their coupling changes across sea surface temperature (SST) regimes and whether the result survives independent satellite observing systems. These sensitivities are consistent with theoretical and observational work on tropical anvil temperature, high-cloud altitude, and longwave cloud feedback (Hartmann & Larson, 2002; Zelinka & Hartmann, 2010; Seeley et al., 2019), while the magnitude and sign of iris-related cloud-area feedback remain debated (Mauritsen & Stevens, 2015; Lindzen & Choi, 2022).

Using approximately the first year of Aura Microwave Limb Sounder (MLS) observations, Su et al. (2006a) showed that upper-tropospheric water vapor ($H_2O$) and cloud ice over tropical oceans were tightly linked and that cloud ice increased rapidly once SST exceeded about 300 K. They interpreted these results as evidence for enhanced convective moistening in the warmest deep-convective regimes. The central question for the present study is not whether the same annual-mean maps can be reproduced with more years, but whether the physical relationship identified from one year is persistent, objectively state dependent, and observable with an independent cloud-ice sensor.

These processes are also relevant to the iris literature, but the present manuscript should not be read as a direct test of the full iris hypothesis. The original iris mechanism concerns changes in cirrus/anvil area per unit convective activity and the resulting longwave and shortwave radiative response (Lindzen et al., 2001; Lindzen & Choi, 2022). By contrast, the primary quantities analyzed here are $H_2O$, cloud ice water content (IWC), ice water path (IWP), and all-cloud top-of-atmosphere (TOA) cloud radiative effect (CRE) binned by SST. MLS is uniquely valuable for observing the upper-tropospheric moisture and ice-condensate branch of deep convection, but it is not, by itself, a measurement of anvil fractional area.

This distinction is important because non-normalized IWP increasing with SST primarily indicates enhanced moistening and condensate-supply capacity in warmer convective regimes. Such behavior can be consistent with iris-type interpretations if anvil or cirrus area per unit convective core decreases after normalization, but that normalization requires an explicit convective-area, convective-core, precipitation-efficiency, or high-cloud-area metric. Previous studies have shown that inferred iris slopes depend on cloud definition, sampling, warm-tail treatment, convective normalization, and the compensation between longwave (LW) and shortwave (SW) CRE (Lin et al., 2002; Hartmann & Michelsen, 2002; Su et al., 2008, 2010; Rondanelli & Lindzen, 2010; Choi et al., 2017; Li et al., 2019).

We therefore adopt a narrower and more defensible goal. We ask whether the moisture and condensate reservoirs associated with tropical deep convection display persistent, SST-dependent behavior over two decades, whether the key cloud-ice transition appears in an independent CALIPSO-A cloud-ice record, and how the condensate branch is related to all-cloud CRE and precipitation. This approach constrains one important part of the iris problem, but it does not estimate the normalized anvil-area feedback itself.

The availability of a two-decade Aura MLS water vapor and ice cloud record, together with a nearly overlapping Cloud-Aerosol Lidar with Orthogonal Polarization (CALIOP)/Cloud-Aerosol Lidar and Infrared Pathfinder Satellite Observations (CALIPSO) ice cloud record, now enables a cross-sensor test that was not possible in the one-year study of Su et al. (2006a). MLS provides long-term vertically resolved microwave limb-sounding observations of $H_2O$ and cloud ice. CALIOP provides higher-resolution active-lidar sampling of cloud vertical structure and is especially sensitive to optically thin ice clouds and deep-convective cloud tops, although it can be attenuated in optically thick clouds. The two observing systems therefore have different viewing geometries, sensitivities, and retrieval assumptions. Agreement between them is a stronger test of the physical relationship than repeating the original MLS-only analysis with a longer average.

In the present study, MLS provides a long 2004–2024 record of $H_2O$, IWC, and IWP, while CALIPSO-A/CALIOP provides an independent active-lidar test of the cloud-ice spatial pattern, the SST-dependent IWP transition, and normalized IWP variability over its 2006–2023 record. MERRA-2 $H_2O$ fields are used only to place a water-vapor column on the CALIPSO grid; they should not be interpreted as a CALIOP water-vapor retrieval. The paper advances beyond Su et al. (2006a) in five ways: it treats the 300 K transition as an objectively estimated breakpoint rather than a prescribed plotting boundary, examines monthly variability rather than only a one-year climatological mean, tests the cloud-ice result with an independent active-lidar ice-cloud product, connects the condensate response to CERES LW, SW, and net CRE, and adds a precipitation-based convective-control diagnostic.

## 2. Data and Methods

### 2.1 Aura MLS $H_2O$, cloud ice, and temperature fields

We use Aura MLS Version 5 (V5) upper-tropospheric $H_2O$, IWC, IWP, and temperature from August 2004 through December 2024. The Aura MLS instrument is described by Waters et al. (2006), and validation and multiyear evaluation of the MLS $H_2O$ and cloud-ice products are given by Read et al. (2007), Wu et al. (2006), Jiang et al. (2010), Livesey et al. (2021), Wang et al. (2021), and the MLS V5 data-quality documentation (MLS Science Team, 2026).

The working gridded product uses 5° longitude by 4° latitude cells, yielding 72 longitudes and 41 latitudes from 2.5° to 357.5° longitude and 80°S to 80°N. Monthly means are first formed from the available MLS observations, and the climatological mean is then computed over all months in the 2004–2024 period. Vertical integrations use a trapezoidal pressure integration and restrict the temperature range to 180–260 K to avoid unphysical or poorly sampled values in the upper-tropospheric integration.

Vertical integrations are calculated using the hydrostatic relation; for example:

$$\mathrm{IWP} = \int_{p_{\mathrm{top}}}^{p_{\mathrm{bottom}}} \mathrm{IWC}\,(p)\,\frac{R_d T(p)}{g_0 p}\,dp \qquad (1)$$

$$C_{H_2O} = \frac{1}{g_0}\int_{p_{\mathrm{top}}}^{p_{\mathrm{bottom}}} q\,(p)\,dp \qquad (2)$$

where $R_d$ is the gas constant for dry air, $g_0$ is gravitational acceleration, $p$ is pressure, $T(p)$ is temperature, $q(p)$ is the $H_2O$ mass mixing ratio, and $C_{H_2O}$ is the vertically integrated $H_2O$ column.

**2.2 CALIPSO active-lidar cloud-ice observations**

As an independent check on the MLS cloud-ice relationships, we use monthly gridded CALIPSO-A/CALIOP ice-cloud fields from June 2006 through June 2023, excluding February 2016 for which no monthly file exists. For the time series, November 2022 was masked due to its anomalous data. Here, "A" denotes the combined day-plus-night CALIPSO sampling; it is distinct from the all-sky IWC averaging convention used below for IWP. We use the CALIPSO-A files, which contain combined day-plus-night monthly statistics, rather than daytime-only or nighttime-only files. This choice provides the closest match to MLS monthly means, which are not restricted to one local-time branch. The CALIPSO mission and CALIOP processing are described by Winker et al. (2009). The analysis follows the CALIOP Level 3 Ice Cloud product framework described by Winker et al. (2024) for $IWC_{\text{in-cloud}}$. We define $IWC_{\text{all-sky}}$ as the product of $IWC_{\text{in-cloud}}$ and ice-cloud fraction, but the sample-count denominator differs slightly: we also include totally attenuated samples and lidar surface/subsurface samples in the denominator. The product is derived from the CALIOP Version 4.20 Level 2 5-km Cloud Profile product and reports global monthly histograms of ice-cloud extinction and IWC on a uniform 2.5° longitude by 2.0° latitude grid and 120 m vertical grid. Because the CALIPSO time series in this manuscript extends to June 2023, the exact processed monthly CALIOP-gridded fields used here are treated as part of the reproducibility package for this study.

The CALIOP IWC retrieval is based on ice-cloud extinction and a temperature-dependent parameterization derived from in situ aircraft observations (Heymsfield et al., 2014; Winker et al., 2024). The CALIOP Level 3 product includes ancillary atmospheric state information from MERRA-2, including pressure and temperature. We use the native MERRA-2 pressure associated with the CALIOP vertical grid for vertical integration, rather than an empirical altitude-pressure conversion. This choice keeps the CALIPSO-A IWP calculation internally consistent with the pressure coordinates used for MERRA-2 $H_2O$ and avoids introducing an additional altitude-to-pressure approximation into the cross-sensor comparison.

Two CALIPSO-A IWP definitions are used, both calculated from the combined day-plus-night all-sky IWC fields. The upper-column IWP integrates IWC over 147–215 hPa, closely matching the upper-tropospheric layer emphasized in the MLS comparison. A second total-column IWP integrates IWC over 147–1020 hPa, using the broader vertical coverage of CALIOP to test whether the SST-dependent behavior is confined to the upper-tropospheric layer or persists when lower-level ice is included. Because CALIOP does not measure water vapor, the $H_2O$ columns shown in the CALIPSO-grid analyses are derived from MERRA-2 relative humidity, temperature, and pressure fields. Saturation vapor pressure over ice is computed using Eq. 7 of Murphy and Koop (2005), which is valid over a wider temperature range. The upper-column $H_2O$ field used in Figure 1 and the upper row of Figure 3 is integrated between 147 and 316 hPa, encompassing the 147–215 hPa upper-IWP layer, while the broader $H_2O$ field used in the lower row of Figure 3 is integrated over 147–1020 hPa, consistent with the corresponding total-column CALIPSO-A IWP analysis.

The CALIPSO-A comparison is used primarily as a test of spatial structure, vertical-layer sensitivity, SST-breakpoint consistency, and normalized temporal variability. We do not require exact agreement in absolute IWP magnitude between MLS and CALIPSO-A because microwave limb sounding and nadir lidar sample different parts of the cloud-ice distribution. CALIOP has high sensitivity to optically thin ice clouds, while the lidar signal can be attenuated in optically thick deep-convective columns; MLS has coarser horizontal resolution and different sensitivity to ice particle size and mass. Cross-sensor agreement in patterns and breakpoints is therefore more meaningful here than one-to-one IWP amplitude agreement.

### 2.3 Auxiliary datasets, greenhouse parameter, radiative fluxes, and precipitation

SST is taken from the NOAA Optimum Interpolation SST (OISST) framework, using the updated monthly OISST V2.1 product (Reynolds et al., 2002; Huang et al., 2021). MERRA-2 provides the relative humidity, temperature, and pressure fields used for the CALIPSO-grid $H_2O$ and CALIOP vertical-integration calculations (Gelaro et al., 2017; Winker et al., 2024). The greenhouse-parameter calculation uses clear-sky outgoing longwave radiation (OLRclr) from the Aqua/AIRS Level 3 monthly standard physical retrieval product (Aumann et al., 2003). Following Raval and Ramanathan (1989) and Su et al. (2006a), the clear-sky greenhouse parameter is defined as:

$$g = \frac{\sigma T_s^4 - \mathrm{OLR_{clr}}}{\sigma T_s^4} \quad (3)$$

where $T_s$ is SST and $\sigma$ is the Stefan-Boltzmann constant.

TOA cloud radiative effects are calculated from CERES Energy Balanced and Filled (EBAF) TOA Edition 4.2.1 monthly all-sky and clear-sky fluxes for August 2004 through December 2024 (Charlock & Ramanathan, 1985; Lacagnina & Selten, 2013; Loeb et al., 2018). We use the standard clear-sky minus all-sky sign convention for the longwave component and define the shortwave component so that enhanced cloud reflection is negative. With OLR denoting outgoing longwave radiation and RSW denoting reflected shortwave radiation at the top of the atmosphere:

$$LW\ CRE = OLR_{clear} - OLR_{all} \qquad (4)$$

$$SW\ CRE = RSW_{clear} - RSW_{all} \qquad (5)$$

$$Net\ CRE = LW\ CRE + SW\ CRE \qquad (6)$$

Positive LW CRE indicates that clouds reduce outgoing longwave radiation relative to clear-sky conditions. Negative SW CRE indicates enhanced solar reflection by clouds, and net CRE is the sum of these two competing effects. For the binned CRE-SST curves, ocean grid-cell monthly values are grouped into 0.5 K SST intervals; bins with fewer than 1000 grid-month samples are excluded from the binned mean. A random subsample is plotted for readability, while the binned means use all valid data.

The state-dependent spatial relationships in Figures 2 and 3 are evaluated over oceanic grid cells between 30°S and 30°N, following Su et al. (2006a). For the time-series, CRE, and precipitation diagnostics in Figures 4–6, the tropical comparison domain is the narrower band between 15°S and 15°N, shown in Figure 1 by dashed black lines. The western Pacific domain used in the original iris analysis, shown by the solid black box, covers 30°S–30°N and 130°E–170°W (equivalently 130°E–190°E). Area means are weighted by cos(latitude). For the normalized time series in Figure 4, MLS and CALIPSO-A IWP are both integrated over 147–215 hPa, the upper-column layer emphasized in the cross-sensor comparison, and the MLS series uses the quality-screened IWC record. CALIPSO-A February 2016 is missing, and November 2022 is excluded as a quality outlier; these months are retained as gaps rather than interpolated.

Consistent with the methodological debate above, precipitation normalization is not applied to the primary MLS $H_2O$ or IWP fields. Normalization by precipitation or convective area is appropriate for a direct cloud-area iris test, but it can obscure the physical meaning of MLS condensate mass when precipitation, convective detrainment, and stratiform cloud fraction are not proportional. As a complementary convective-control diagnostic, we relate MLS IWP to GPM IMERG Final Run monthly precipitation (GPM_3IMERGM, Version 07B; Huffman et al., 2023).

The precipitation field is regridded from 0.1° resolution to the MLS 5° × 4° grid, and the MLS IWP used in this diagnostic is integrated over 147–215 hPa using the same temperature screening and hydrostatic integration described above. This upper-column layer targets the convective detrainment region where the MLS cloud-ice signal is strongest and is the most direct layer for comparison with the CALIPSO-A upper-column analysis. This comparison tests whether upper-tropospheric IWP covaries with convective rain rate; it is not used as a normalization of cloud area by convective area.

## 2.4 Regression and breakpoint analysis

The $H_2O$-IWP relationship is evaluated in log-log space, following the original Su et al. (2006a) framework. The generic fitting form is:

$$\ln(H_2O) = a\ln(\mathrm{IWP}) + b \qquad (7)$$

where $a$ is the fitted slope and $b$ is the fitted intercept.

For the SST-dependent relationships, we use a continuous one-knot segmented regression:

$$y = \begin{cases} \beta_0 + \beta_1(T - T_b) + \varepsilon, & T \le T_b, \\ \beta_0 + (\beta_1 + \beta_2)(T - T_b) + \varepsilon, & T > T_b \end{cases} \qquad (8)$$

where $y$ is the fitted response variable, $\beta_0$ is the fitted value at the breakpoint, $T$ is SST, and $T_b$ is the fitted breakpoint. Breakpoints are estimated by residual-sum-of-squares minimization over a grid of candidates; a Davies-type test for a change in slope (Davies, 1987) is used to determine breakpoint existence, and 95% confidence intervals are obtained by profile-likelihood. $H_2O$ and ice fits are performed in log space when appropriate; the clear-sky greenhouse parameter is fitted in linear space.

## 2.5 MLS V5 data-quality treatment for the 2024 extension

This study uses a single MLS V5 processing stream throughout, so all figures and regression values are internally consistent with the same retrieval version. After May 2024, the MLS 190 GHz subsystem began duty-cycled operation, resulting in degrading data quality. Anomalous IWC retrievals throughout 2023 and 2024 were removed using percentile-based outlier masking by year and pressure level. Negative IWC values were masked throughout the entire 2004 to 2024 record. The screened IWC dataset is used only for Figure 4 as time series are sensitive to retrieval quality. Because small multiyear $H_2O$ trends can be sensitive to instrument and retrieval stability (Livesey et al., 2021), the present manuscript emphasizes climatological relationships, regime structure, cross-sensor consistency, and monthly variability rather than interpreting small linear trends.

## 3. Results

### 3.1 Two-decade climatological structure of upper-tropospheric $H_2O$ and cloud ice

Figure 1 compares the 2004–2024 MLS climatological mean with the CALIPSO-grid 2006–2023 analysis. The CALIPSO-grid $H_2O$ map is MERRA-2 $H_2O$ integrated between 147 and 316 hPa, not a CALIOP retrieval, while the CALIPSO-A cloud-ice map is CALIOP-derived IWP integrated over 147–215 hPa using native MERRA-2 pressure and combined day-plus-night sampling. Both the MLS and CALIPSO-grid analyses place the strongest moisture and cloud-ice maxima over the Indo-Pacific warm pool and Maritime Continent, with additional cloud-ice maxima over tropical Africa and South America. The spatial agreement is important because MLS samples the limb with relatively coarse horizontal resolution, while CALIOP provides an independent active-lidar view of cloud vertical structure. The agreement indicates that the dominant convective moistening and cloud-ice pattern is not specific to one retrieval system.

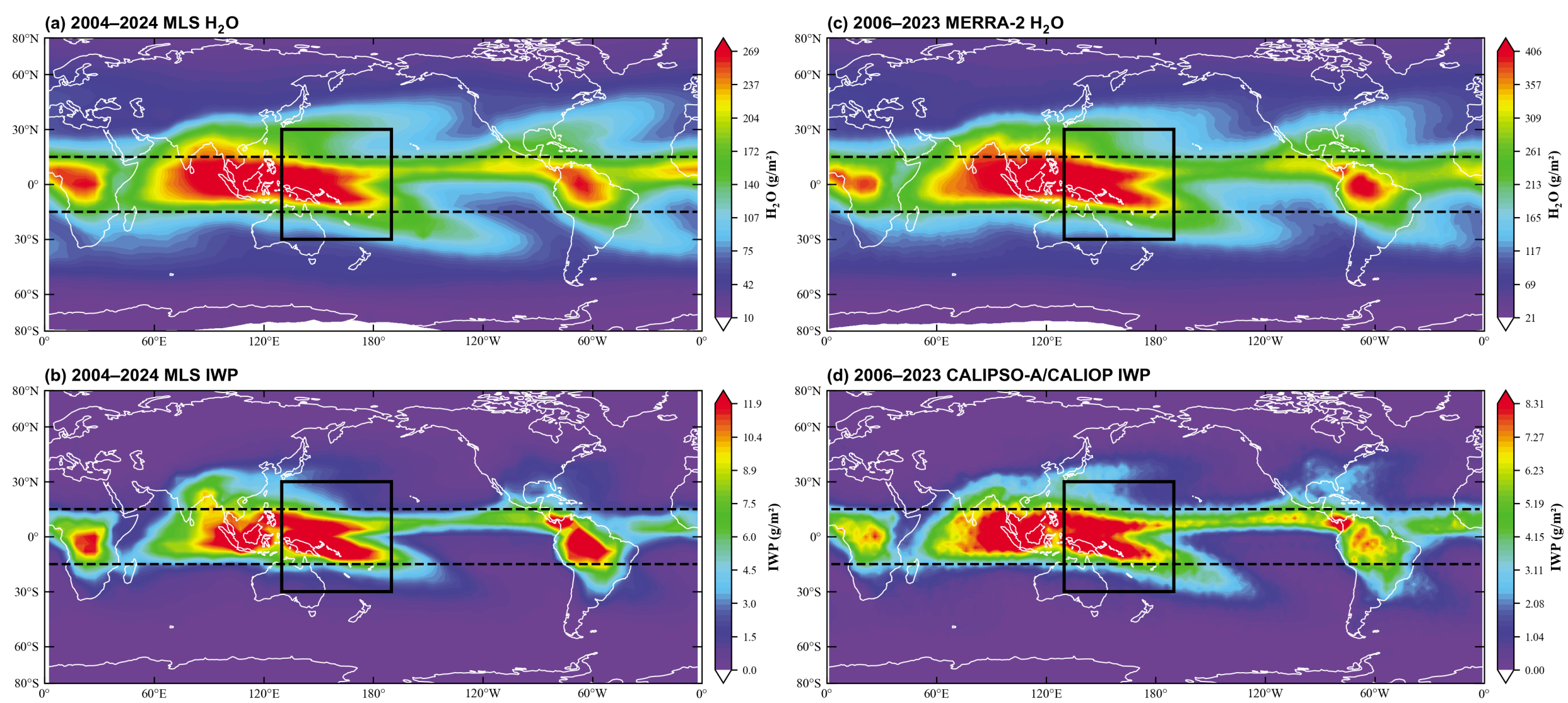

**Figure 1.** Climatological mean $H_2O$ and IWP from CALIPSO-grid and MLS analyses. The left panels show the 2004–2024 MLS $H_2O$ (a) and IWP columns (b) for the corresponding upper-tropospheric layers. Dashed black lines indicate 15°S and 15°N, and the solid box denotes the western Pacific Lindzen domain. The upper-right panel shows the 2006–2023 MERRA-2 $H_2O$ column (c) on the CALIPSO grid, integrated between 147 and 316 hPa. The lower-right panel shows 2006–2023 CALIPSO-A/CALIOP IWP (d) integrated over 147–215 hPa using combined day-plus-night CALIPSO-A sampling and native MERRA-2 pressure.

The dashed black lines in Figure 1 show the ±15° tropical band used for the tropical-mean time series, whereas the solid box shows the western Pacific domain used in the original iris analysis. The $H_2O$ field is broader than the cloud-ice field in both the MLS and CALIPSO-grid analyses, as expected: water vapor can be transported and mixed after convective detrainment, whereas large cloud-ice values remain more tightly tied to active deep-convective regions. The agreement between MLS and CALIPSO-A/CALIOP in the location of the dominant cloud-ice maxima provides the first independent check on the spatial component of the MLS result.

### 3.2 Coupled $H_2O$-cloud-ice relationship and SST-state dependence

Figure 2a quantifies the $H_2O$-cloud-ice relationship over tropical-ocean grid cells between 30°S and 30°N in the 2004–2024 mean. The fitted relationship is:

$$\ln(H_2O) = 0.365\ln(\mathrm{IWP}) + 4.62, \qquad r = 0.96. \qquad (9)$$

This confirms that the strong $H_2O$-ice relationship identified in Su et al. (2006a) based on the first year of Aura MLS observations persists in the two-decade climatology. The slope is less than unity, indicating that upper-tropospheric water vapor increases more slowly than cloud ice in log space, but the high correlation demonstrates that the two quantities are controlled by a common deep-convective regime.

In Figure 2, the color of each point represents the number of grid boxes contributing to that bin, so the densest parts of the tropical-ocean distribution are visually separated from isolated outliers. Red dashed lines mark the estimated breakpoints, and red shading marks the corresponding 95% confidence interval. Figures 2b–2d show that the relationships with SST are strongly state dependent. Cloud ice remains relatively small and scattered at lower tropical SSTs, but increases sharply once SST exceeds the fitted breakpoint of 299.15 K. The corresponding $H_2O$-SST relationship has a breakpoint of 299.45 K and a much steeper post-breakpoint slope.

The fitted sensitivities of $\ln(H_2O)$ to SST are:

$$\frac{d\ln(H_2O)}{dT} = 0.068\ \mathrm{K}^{-1} \quad (T < T_b), \quad \frac{d\ln(H_2O)}{dT} = 0.246\ \mathrm{K}^{-1} \quad (T > T_b) \qquad (10)$$

This change in slope is consistent with the idea that upper-tropospheric moistening is amplified in the warmest deep-convective regimes.

The clear-sky greenhouse parameter also shows a state-dependent response, with a fitted breakpoint near 301.45 K. The higher breakpoint for g than for cloud ice or $H_2O$ likely reflects the fact that the clear-sky greenhouse parameter integrates the radiative influence of temperature and water vapor over a broader atmospheric column, rather than responding only to the local upper-tropospheric condensate field. Together, the three breakpoints show that the warm-pool transition is not a visually imposed 300 K threshold; it emerges from the two-decade gridded relationships.

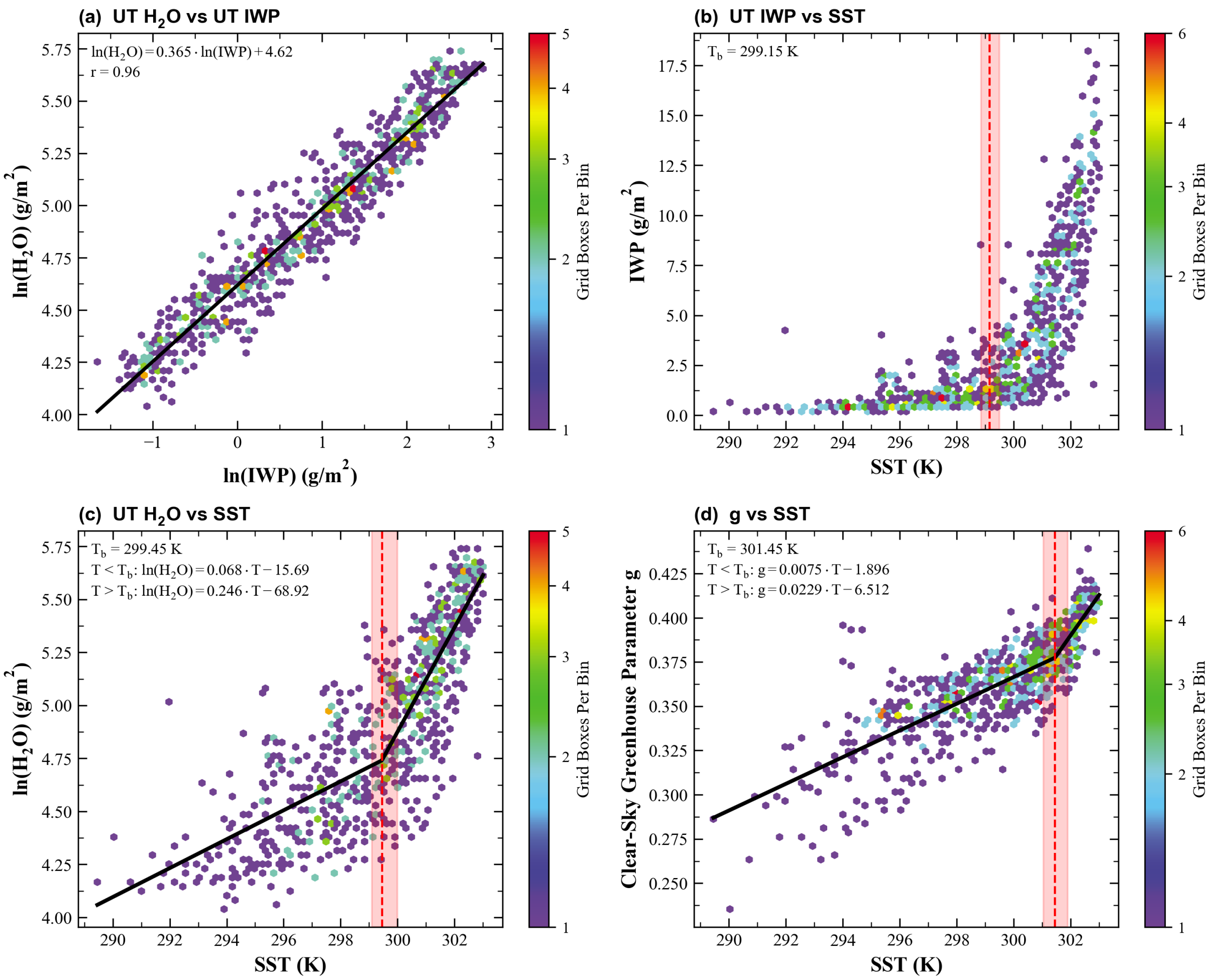

**Figure 2.** State-dependent relationships over tropical-ocean grid cells between 30°S and 30°N in the 2004–2024 climatological mean. (a) ln($H_2O$) versus ln(IWP), with fitted slope 0.365, intercept 4.62, and correlation coefficient 0.96. (b) IWP versus SST, with fitted SST breakpoint 299.15 K (95% CI: 298.85–299.50 K). (c) ln($H_2O$) versus SST, with fitted breakpoint 299.45 K (95% CI: 299.10–300.00 K). (d) Clear-sky greenhouse parameter g versus SST, with fitted breakpoint 301.45 K (95% CI: 301.05–301.90 K). Red dashed lines indicate fitted breakpoints, red shading indicates the 95% confidence interval, and point colors indicate the number of grid boxes per bin.

### 3.3 Independent CALIPSO check on state-dependent cloud ice

CALIPSO-A provides a second observational test of whether the MLS cloud-ice relationships are robust across sensors and vertical definitions. Figure 3 repeats the $H_2O$-IWP and SST-dependent analyses using the CALIPSO-grid climatology over oceanic grid cells between 30°S and 30°N for two IWP integrations. The upper row uses CALIPSO-A IWP integrated over 147–215 hPa and MERRA-2 $H_2O$ integrated over 147–316 hPa on the CALIPSO grid, emphasizing the upper-tropospheric layer most directly comparable to the MLS cloud-ice analysis. In this layer, the ln($H_2O$)-ln(IWP) relationship remains strong, with a fitted slope of 0.357 and correlation coefficient r = 0.94. The upper-column CALIPSO-A IWP-SST relationship has a fitted breakpoint of 299.35 K. The corresponding MERRA-2 $H_2O$-SST relationship on the CALIPSO grid has a fitted breakpoint of 299.60 K.

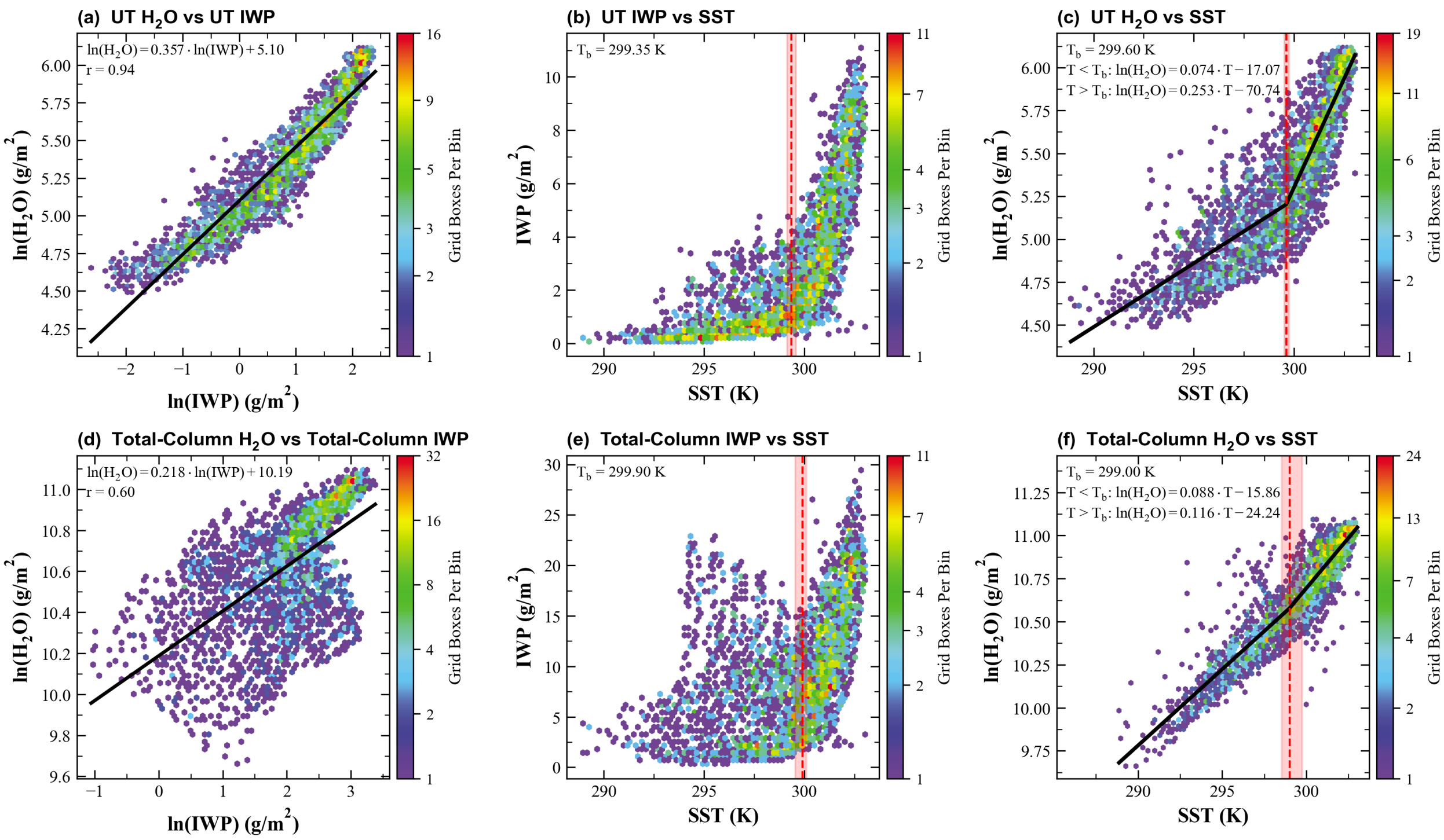

**Figure 3.** CALIPSO-A-grid state-dependent relationships over tropical-ocean grid cells between 30°S and 30°N for 2006–2023. The upper row uses CALIPSO-A IWP integrated over 147–215 hPa and MERRA-2 $H_2O$ on the CALIPSO grid integrated over 147–316 hPa. The lower row uses total-column CALIPSO-A IWP and MERRA-2 $H_2O$ on the CALIPSO grid integrated over 147–1020 hPa. For the upper row, $\ln(H_2O) = 0.357 \ln(IWP) + 5.10$ with r = 0.94, the IWP-SST breakpoint is 299.35 K (95% CI: 299.15–299.60 K), and the $H_2O$-SST breakpoint is 299.60 K (95% CI: 299.50–299.75 K). For the lower row, $\ln(H_2O) = 0.218 \ln(IWP) + 10.19$ with r = 0.60, the total-column IWP-SST breakpoint is 299.90 K (95% CI: 299.55–300.10 K), and the corresponding MERRA-2 $H_2O$-SST breakpoint is 299.00 K (95% CI: 298.55–299.75 K). Red dashed lines indicate fitted breakpoints, red shading indicates the 95% confidence interval, and point colors indicate the number of grid boxes per bin.

The lower row of Figure 3 uses the broader CALIPSO-A total-column IWP and MERRA-2 $H_2O$ integrated over 147–1020 hPa. The total-column $\ln(H_2O)$-$\ln(IWP)$ relationship is weaker than the upper-column relationship, with slope 0.218 and r = 0.60, because the broader column includes lower and middle tropospheric ice and water vapor that are less directly tied to upper-tropospheric convective detrainment and water-vapor outflow. Nevertheless, the total-column IWP-SST breakpoint remains close to the MLS and upper-column CALIPSO-A values, at 299.90 K. The corresponding MERRA-2 $H_2O$-SST breakpoint is 299.00 K. This vertical-layer comparison strengthens the main result: the warm-regime transition is most tightly expressed in upper-tropospheric cloud ice but remains visible even when the broader CALIOP ice column is included.

### 3.4 Time evolution in the tropical mean and western Pacific box

The spatial relationships in Figures 1–3 establish climatological and state-dependent coupling among $H_2O$, cloud ice, IWP, and SST. Figure 4 compares normalized MLS and CALIPSO-A upper-column IWP integrated over 147–215 hPa for the western Pacific Lindzen box and the 15°S–15°N tropical mean using each sensor's available record: MLS from August 2004 through December 2024 and CALIPSO-A from June 2006 through June 2023. Showing the full MLS record preserves the two-decade context while retaining the independent CALIPSO-A comparison during the active CALIPSO mission. The normalization removes the absolute-amplitude offset expected from different instrument sensitivities and focuses the comparison on temporal variability.

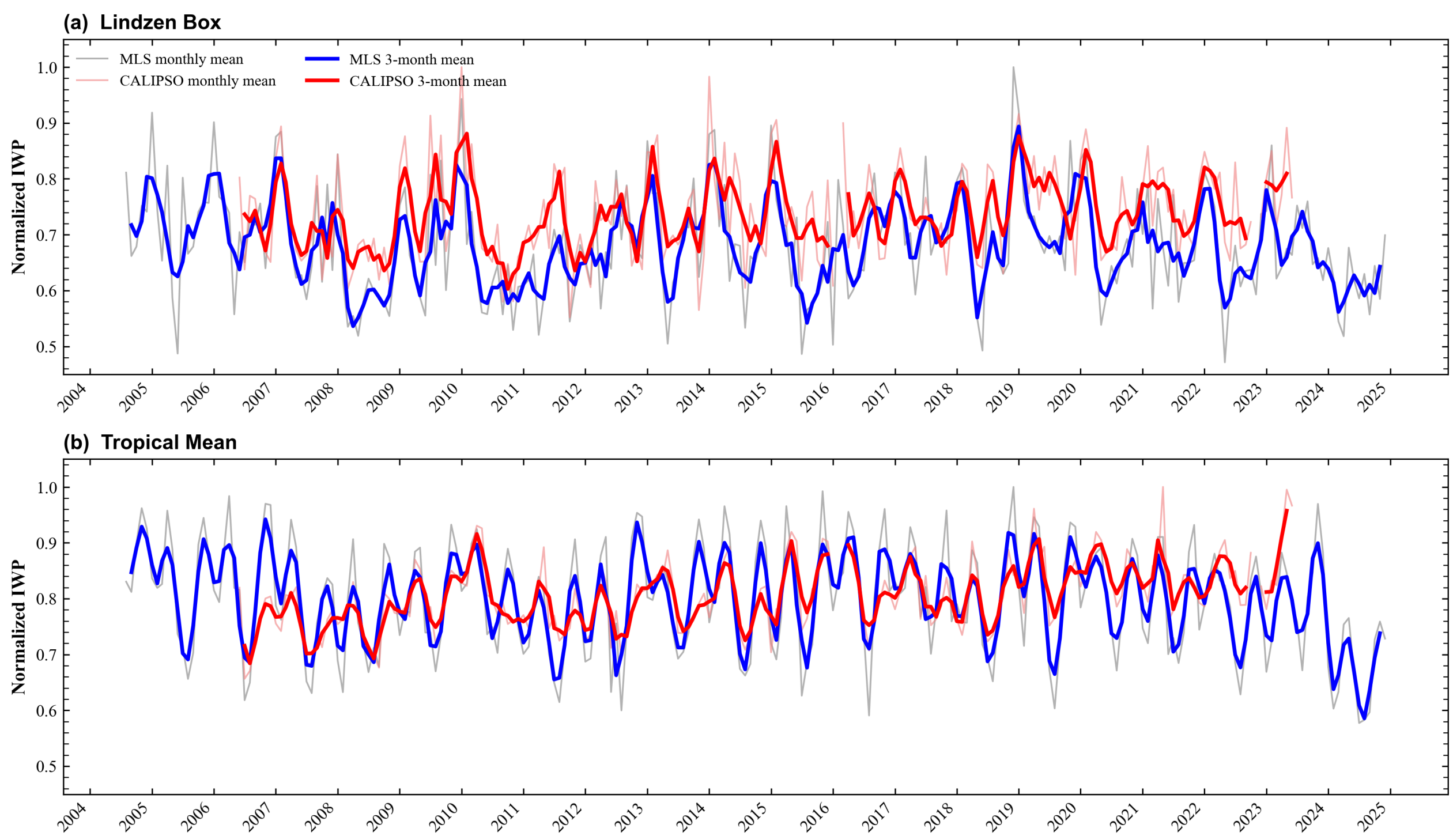

**Figure 4.** Normalized IWP time series from MLS and CALIPSO-A in (a) the western Pacific Lindzen box and (b) the tropical mean within 15°S–15°N. MLS spans August 2004–December 2024, and CALIPSO-A spans June 2006–June 2023. Gray and light red curves show monthly means for MLS and CALIPSO-A, respectively; blue and red curves show twice-applied centered 3-month means for MLS and CALIPSO-A, respectively. The breaks near February 2016 and November 2022 are consistent with the missing monthly file for February 2016 and the masked data during November 2022. The comparison emphasizes temporal co-variability and persistence rather than absolute IWP amplitude agreement.

In the western Pacific box, both MLS and CALIPSO-A show persistent month-to-month and interannual IWP variability, consistent with local warm-pool convection, ENSO, the Madden-Julian Oscillation, and seasonal shifts in convective organization. The tropical-mean series is smoother because averaging over all longitudes damps regional convective extremes. The two sensors do not agree exactly month by month, which is expected from the differences between limb microwave and nadir lidar sampling, but they show broadly comparable variability during the 2006–2023 overlap and preserve a substantial upper-tropospheric IWP reservoir. The extended MLS portion before 2006 and after mid-2023 provides context for the longer MLS record without being interpreted as a cross-sensor comparison. Thus the warm-regime cloud-ice response is not only a feature of climatological maps and scatterplots; it is also expressed in the time-evolving condensate reservoir observed by two independent satellite systems.

### 3.5 Cloud-radiative response and precipitation as cloud-controlling diagnostics

Figures 1–4 constrain the moisture and condensate reservoirs associated with tropical deep convection. To extend the analysis toward the radiative branch of the problem, Figure 5 applies a CERES-based cloud-radiative diagnostic in which TOA LW, SW, and net CRE are binned by SST over both the western Pacific Lindzen box and the tropical band.

In both regions, LW CRE increases sharply in the warm SST regime, broadly consistent with the increase in IWP identified by MLS and CALIPSO. At the same time, SW CRE becomes more negative at high SSTs, indicating stronger solar reflection by clouds. As a result, the net CRE response is much smaller than the LW response alone. In the Lindzen box, the binned net CRE increases from about −20 W $m^{-2}$ at cooler tropical SSTs toward weakly negative values in the warmest bins. In the tropical mean, the net CRE also becomes less negative with increasing SST, although the relationship reflects a mixture of cloud regimes across the full 15°S–15°N band. These results are consistent with the LW and SW CRE relationships with SST reported by Lacagnina and Selten (2013), accounting for their opposite SW CRE sign convention.

This compensation is important for interpreting the iris hypothesis. A LW-only diagnostic would suggest a stronger cloud greenhouse effect in the warm-pool regime, but the CERES SW and net CRE panels show that enhanced shortwave reflection offsets part of that longwave effect. Therefore, the net radiative behavior of the warm-pool cloud response cannot be inferred from IWP or LW CRE alone.

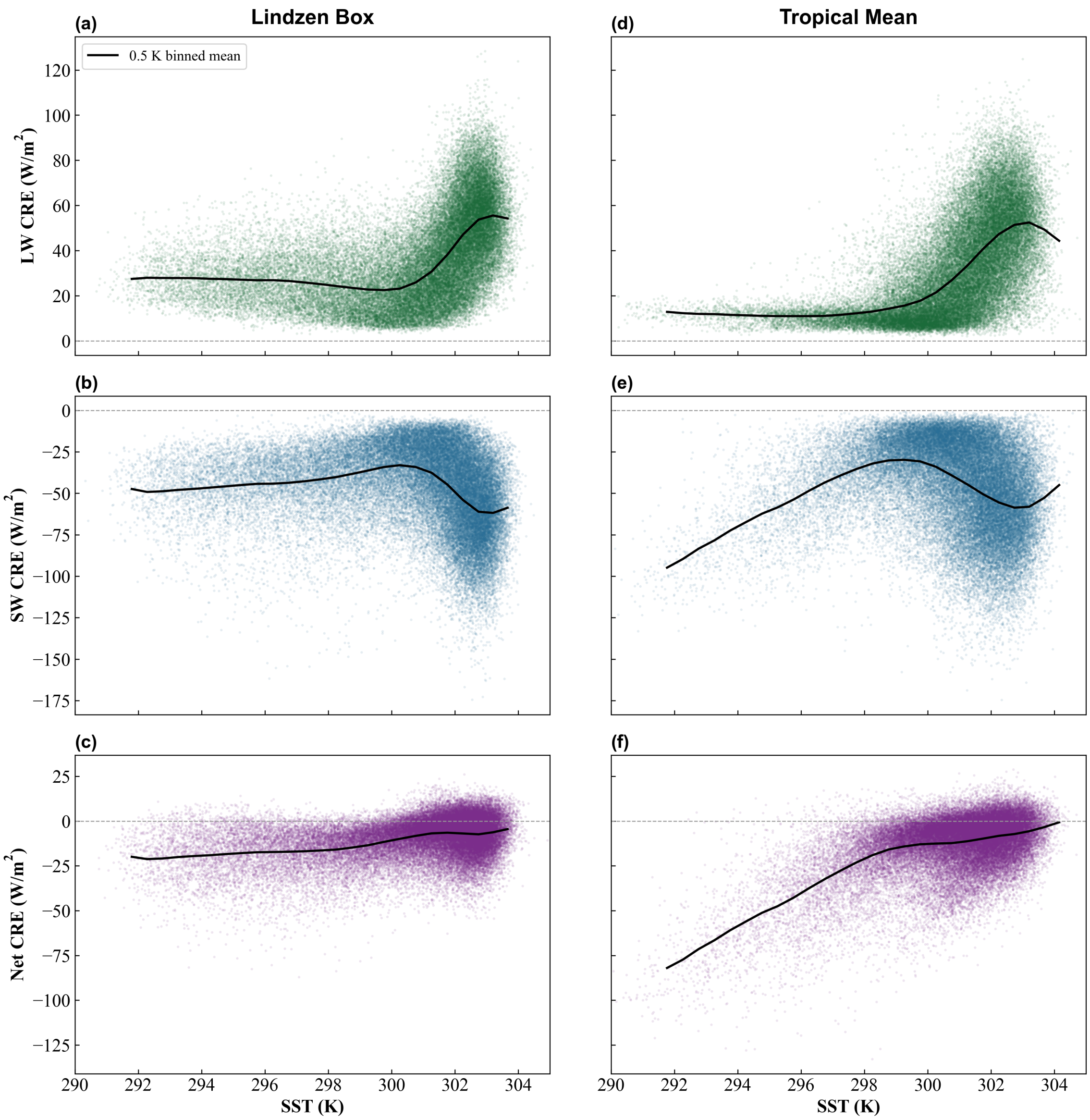

**Figure 5.** CERES top-of-atmosphere CRE versus SST over ocean grid cells for August 2004–December 2024. The left column shows the western Pacific Lindzen box, and the right column shows the tropical band within 15°S–15°N. Rows show LW CRE, SW CRE, and net CRE, respectively. CREs are computed from CERES EBAF TOA Edition 4.2.1 monthly all-sky and clear-sky fluxes. SST is from NOAA OISST V2.1 regridded to the CERES 1° × 1° grid. Points show a random subsample of valid ocean grid-months for readability, and black curves show 0.5 K binned means after excluding SST bins with fewer than 1000 samples. With the sign convention used here, positive LW CRE indicates cloud longwave warming, negative SW CRE indicates cloud shortwave cooling, and net CRE equals LW CRE plus SW CRE.

Figure 6 provides a complementary convective-control diagnostic by relating MLS upper-column IWP to GPM IMERG precipitation. IWP increases with precipitation in both the tropical band and the Lindzen box, with correlations of $r = 0.653$ and $r = 0.679$, respectively. The fitted relationships are $IWP = 19.830P + 2.631$ for the tropical band and $IWP = 24.087P + 1.755$ for the Lindzen box, where P is precipitation in mm hr⁻¹. The larger slope in the Lindzen box is consistent with the concentration of deep-convective detrainment in the western Pacific warm pool. This analysis supports the interpretation that the observed IWP variability is physically connected to convective rain rate, not only to the spatial sampling of warm SSTs.

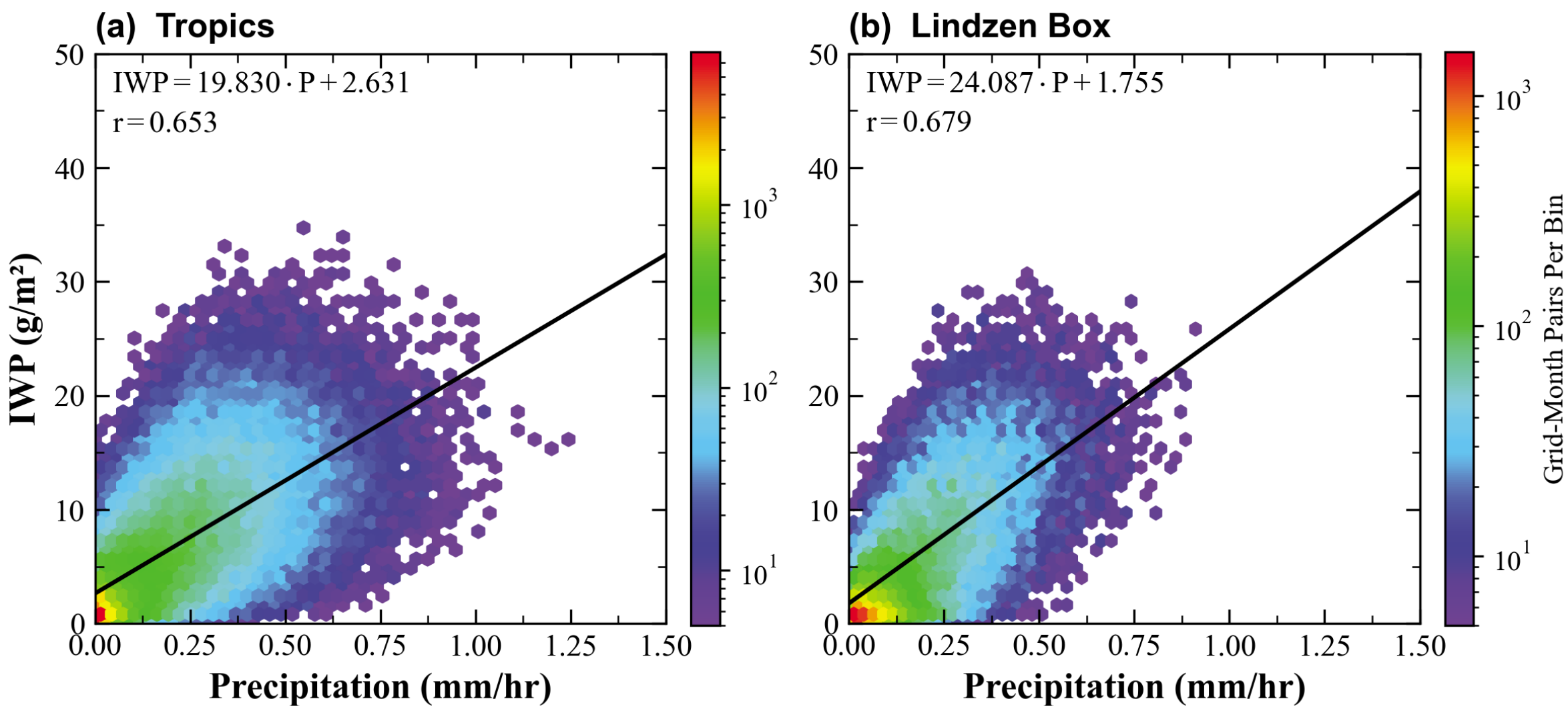

**Figure 6.** MLS IWP versus GPM IMERG precipitation over ocean grid cells for August 2004–December 2024. (a) Tropical band within 15°S–15°N. (b) Western Pacific Lindzen box. MLS IWP is integrated over 147–215 hPa using the hydrostatic relation and the same temperature screening used in the MLS time-series analysis. GPM IMERG Final Run monthly precipitation (Version 07B) is regridded to the MLS grid. Hexbin colors indicate the number of grid-month pairs per bin, and black lines show least-squares fits. The positive correlations indicate that upper-tropospheric IWP increases with convective rain rate in both domains.

## 4. Discussion and Conclusions

The main contribution of this study is a cross-sensor and cross-diagnostic characterization of the warm-pool transition in convective moistening, rather than a simple extension of the original MLS climatology. By combining a two-decade MLS V5 record with CALIPSO-A/CALIOP ice-cloud retrievals, MERRA-2 $H_2O$ on the CALIPSO grid, objective breakpoint fitting, CERES LW, SW, and net CRE, and GPM precipitation, the analysis tests whether the moisture-condensate relationship is persistent, reproducible across observing systems, and physically tied to convection and radiation. The core features identified from MLS, tight $H_2O$-cloud-ice coupling, rapid increase of cloud ice in the warmest tropical SST regime, and enhanced upper-tropospheric moistening associated with deep convection, are preserved in the 2004–2024 climatology and are independently supported by CALIPSO-A. The objective breakpoints near 299–300 K show that the warm-pool transition is an observed regime change rather than a prescribed plotting boundary.

The physical interpretation is straightforward. Deep tropical convection lofts both water vapor and ice condensate into the upper troposphere. Some condensate is detrained as anvil cloud, some precipitates, and some sublimates or evaporates into the surrounding environment. The resulting upper-tropospheric moistening strengthens longwave trapping, while the cloud ice itself contributes to both longwave and shortwave cloud-radiative effects. Once SST reaches the warm-pool regime near 299–300 K, the probability of deep convection and the magnitude of convective detrainment increase sharply, producing the observed SST-dependent slopes. The positive IWP-precipitation relationship provides an independent dynamical check on this interpretation because precipitation is a more direct indicator of convective activity than SST alone. The stronger CALIPSO relationship for 147–215 hPa IWP than for 147–1020 hPa total-column IWP further supports this interpretation because the upper-column layer is closer to the convective outflow and anvil-detrainment region central to the MLS moisture-condensate response.

The relationship to the iris hypothesis should be stated carefully. The increase of non-normalized IWP with SST is best interpreted as an increase in convective moistening and condensate-supply capacity in the warm-pool regime. This result is not inconsistent with an iris-type mechanism, because the iris effect concerns the area of cirrus or anvil cloud per unit convective activity and the net radiative effect of that area change. Our CERES and GPM analyses provide relevant radiative and convective context, but they do not normalize high-cloud area by convective-core area or directly estimate an anvil-area feedback. Therefore, the present results constrain the moisture-condensate branch of the problem rather than proving or disproving the full iris hypothesis.

This distinction places the studies by Su et al. (2008), Rondanelli and Lindzen (2010), Su et al. (2010), Choi et al. (2017), and Lindzen and Choi (2022) in the appropriate context: they motivate our interpretation of the moisture-condensate pathway, while also clarifying why a complete iris-effect test requires convective normalization and explicit cloud-area diagnostics. The inferred iris slope depends on whether one uses cloud fraction, cirrus area normalized by convective cloud area, precipitation-normalized cirrus amount, radiative flux, or condensate mass. Our analysis avoids re-estimating the original normalized cloud-area slope from MLS alone. Instead, it provides a robust observational benchmark for the part of the chain that MLS and CALIPSO-A observe most directly: whether the warmest convective regimes produce a persistent increase in upper-tropospheric moisture and ice condensate and how that condensate branch relates to precipitation and all-cloud CRE.

The model-comparison perspective of Su et al. (2006b) also remains important. Their comparison of MLS observations with ECMWF analyses and GCM simulations showed that models may capture the qualitative relationship among upper-tropospheric water vapor, cloud ice, IWP, and SST while differing in amplitude. The present study provides a more stringent observational benchmark: models should reproduce not only the spatial collocation of $H_2O$ and cloud ice, but also the segmented SST dependence, the tropical-versus-western-Pacific contrast, the persistence of normalized IWP variability on seasonal-to-interannual time scales, the fact that the warm-regime transition appears in both microwave-limb and combined day-night active-lidar cloud-ice records, the positive IWP-precipitation relationship, and the LW/SW compensation in CERES net CRE. Because MERRA-2 $H_2O$ is used only for the CALIPSO-grid moisture field, model comparisons should treat the CALIPSO-A analysis primarily as an independent IWP and vertical-structure constraint rather than as an independent satellite $H_2O$ retrieval.

The main limitations also help define the contribution of this work. First, long-term trend interpretation of MLS $H_2O$ requires caution, so the present analysis treats the two-decade record primarily as a test of robust relationships and regime behavior. Second, all MLS quantitative values are V5-specific, and the late-2023-2024 extension requires strict screening because duty-cycled 190-GHz operation affects $H_2O$ and cloud-ice retrievals on inactive days. Third, MLS cloud ice cannot resolve very thin cirrus with the same sensitivity as lidar, while CALIOP can be affected by lidar attenuation in optically thick deep convection, narrow-swath sampling, and the temperature-dependent IWC-extinction parameterization. Fourth, we use combined day-plus-night CALIPSO-A fields to match the all-local-time MLS monthly means; nighttime-only CALIOP retrievals may have higher thin-cloud sensitivity, but they would introduce a local-time sampling mismatch for the main comparison. Fifth, the CALIPSO-grid $H_2O$ panels use MERRA-2 $H_2O$ and therefore are not independent satellite water-vapor retrievals. Sixth, CERES CRE is an all-cloud TOA flux diagnostic rather than a high-cloud-only or anvil-only radiative diagnostic. Seventh, GPM precipitation is a useful convective proxy but is not identical to convective cloud area or detrainment rate. Eighth, the current analysis uses monthly and climatological mean relationships rather than event-scale convective life-cycle tracking. These limitations argue for future work combining MLS, CALIPSO-A, high-cloud fraction, precipitation, CERES radiative fluxes, and explicit high-cloud/anvil tracking.

In summary, the 2004–2024 MLS V5 record preserves strong spatial collocation of upper-tropospheric $H_2O$ and cloud ice over tropical deep-convective regions, especially the Indo-Pacific warm pool and Maritime Continent. The log-log $H_2O$-IWP relationship remains tight over tropical-ocean grid cells between 30°S and 30°N, and objective segmented regressions identify warm-regime transitions near 299–300 K. The independent 2006–2023 CALIPSO-A analysis reproduces the dominant cloud-ice maxima and shows that the warm-regime IWP transition appears in both an upper-column 147–215 hPa integration and a broader 147–1020 hPa total-column integration, with the upper-column relationship more tightly coupled to MERRA-2 $H_2O$ on the CALIPSO grid. Normalized MLS and CALIPSO-A IWP time series further show persistent variability in the western Pacific and tropical mean, with cross-sensor consistency during the 2006–2023 common interval and continuation of the MLS cloud-ice record through 2024. CERES shows that the condensate-rich warm regime has stronger LW CRE but also stronger SW cooling, yielding partial compensation in net CRE. GPM IMERG precipitation further shows that MLS IWP increases with rain rate, confirming the convective character of the IWP response. Together, these observations show that convective moistening and condensate production across the tropical warm-pool transition are persistent, cross-sensor features of deep convection with associated radiative and precipitation signatures. They do not by themselves prove or disprove the full iris hypothesis, but they provide a substantially stronger observational constraint on the moisture-condensate branch and its relationship to cloud-radiative behavior.

**Acknowledgments**

We acknowledge the NASA Aura MLS, CALIPSO/CALIOP, MERRA-2, AIRS, CERES, GPM IMERG, and NOAA OISST science and data teams for producing and maintaining the satellite, reanalysis, precipitation, radiative-flux, and SST records that make this analysis possible. This research received no government funding. The research and publication costs were supported by the Jonathan H. Jiang Foundation.

**Availability Statement**

Aura MLS Version 5 $H_2O$, IWC, IWP, and temperature observations are publicly available through the NASA MLS/JPL project and NASA GES DISC data systems. CALIPSO/CALIOP Level 3 Ice Cloud data and related CALIOP cloud profile products are publicly available through NASA ASDC/Earthdata, including the CALIPSO Lidar Level 3 Ice Cloud Data, Standard V2-00 product; the combined day-plus-night CALIPSO-A monthly fields and scripts used in this analysis will be deposited in a public repository before publication and are available from the corresponding author during review. MERRA-2 meteorological fields are publicly available from NASA GES DISC/GMAO. NOAA OISST products are publicly available from NOAA NCEI. AIRS Level 3 monthly standard physical retrieval products are publicly available from NASA GES DISC. CERES EBAF TOA Edition 4.2.1 monthly radiative fluxes are available from the NASA CERES project. GPM IMERG Final Run monthly precipitation (GPM_3IMERGM, Version 07B) is available from NASA GES DISC.